\documentclass[12pt]{article}

 \usepackage{amssymb}
 \usepackage{amsmath}
 \usepackage{amsfonts}
 \usepackage{color}
 \usepackage{dcolumn}
 \usepackage{hyperref}
 \numberwithin{equation}{section}
 \newcommand{\be}{\begin{equation}}
 \newcommand{\ee}{\end{equation}}
 \newcommand{\bea}{\begin{eqnarray}}
 \newcommand{\eea}{\end{eqnarray}}
 \newcommand{\nn}{\nonumber}
 \newcommand{\rd}{\partial}

\begin{document}

 \begin{titlepage}
  \thispagestyle{empty}

  \hfill  WITS-CTP-79

  \vspace{2cm}

  \begin{center}
    \font\titlerm=cmr10 scaled\magstep4
    \font\titlei=cmmi10 scaled\magstep4
    \font\titleis=cmmi7 scaled\magstep4
     \centerline{\titlerm  CFT Duals of Black Rings With Higher Derivative Terms}

    \vspace{1.5cm}
    \noindent{{
        Kevin Goldstein\footnote{e-mail:kevin (at) neo.phys.wits.ac.za },
        Hesam Soltanpanahi\footnote{e-mail:hesam.soltanpanahisarabi (at) wits.ac.za }
         }}\\
    \vspace{0.8cm}

  {\it
National Institute for Theoretical Physics, Gauteng\\
School of Physics and Center for Theoretical Physics,\\
 University of the Witwatersrand,\\ WITS 2050, Johannesburg, South Africa}

  \end{center}

  \vskip 2em

  \begin{abstract}
    We study possible CFT duals of supersymmetric five dimensional
    black rings in the presence of supersymmetric higher derivative
    corrections to the ${\cal N}=2$ supergravity action. A Virasoro
    algebra associated to an asymptotic symmetry group of solutions is
    defined by using the Kerr/CFT approach. We find the central charge
    and compute the microscopic entropy which is in precise agreement
    with the macroscopic entropy. Although apparently related to a
    different aspect of the near-horizon geometry and a different
    Virasoro algebra, we find that the c-extremization method leads to
    the same central charge and microscopic entropy computed in the
    Kerr/CFT approach. The relationship between these two point of
    view is clarified by relating the geometry to a self-dual orbifold
    of AdS$_3$.
  \end{abstract}

\end{titlepage}

 \tableofcontents


 \section{Introduction}

 A decade before Maldacena's AdS/CFT conjecture of gravity-gauge
 duality \cite{9711200}, a relationship between gravity on AdS$_3$ and
 the two dimensional conformal group was found by Brown and Henneaux
 \cite{Brown:1986nw}.  They found that appropriate boundary
 conditions, which preserve the asymptotically AdS$_3$ geometry,
 generate two Virasoro algebras. It is not a great leap to relate this
 asymptotic symmetry group to some CFT living on the boundary and
 applying the Brown-Henneaux formalism to asymptotically AdS black
 hole solutions allows one to relate the macroscopic
 Bekenstein-Hawking entropy to the number of states of a boundary CFT
 \cite{9712251, 9812013, 9812056, 0111224}.  One uses the fact that if
 the asymptotic symmetry group is a Virasoro algebra with a particular
 central charge, the Cardy formula relates the central charge to
 entropy of the CFT in the large charge limit. We refer to the entropy
 obtained from the Brown-Henneaux formalism as the microscopic entropy
 but it should be recalled that, in many cases, the existance of a
 microscopic CFT is only hypothesised.

 More recently, a clever extension of the Brown-Henneaux approach to
 the SL(2,${\mathbb R}$)$\times U(1)$ near-horizon symmetry of the
 extremal Kerr black hole, led to the proposal of a Kerr/CFT
 correspondence \cite{0809.4266}.  It was found that one can find an
 asymptotic symmetry group, corresponding to the $U(1)$ part of the isometry,
 with a Virasoro algebra whose central charge correctly accounts for
 the Bekenstein-Hawking entropy via the Cardy formula.

 The Kerr/CFT correspondence has been generalised to, and verified for
 higher dimensional rotating black holes \cite{0810.2620, 0811.2225,
   0811.4177, 0811.4393, 0812.2918, 0812.4440, 0901.0311, 0901.1595},
 Lagrangians with topological terms in four and five dimensions
 \cite{0902.1001} and Lagrangians with higher derivative terms
 \cite{0903.4176}.  Notably, it was shown that in certain theories
 with gravity coupled to matter fields the associated central charge
 coming from the gravitational degrees of freedom is sufficient to
 account for the entropy \cite{0811.4393, 0902.1001}.

 Recently, people have tried to embed the Kerr/CFT approach in string
 theory \cite{1009.5039, 1010.0685, 1101.5136}. In light of this we
 were motivated to study possible CFT duals of five dimensional
 supersymmetric black rings \cite{0110260} in the presence of special
 higher derivative terms which are the supersymmetric completion of a
 mixed gauge-gravitational Chern-Simons term \cite{0611329}.  These
 solutions have SL(2,${\mathbb R}$)$\times U(1)\times$SO(3)
 near-horizon symmetries corresponding to an S$^1$ fibred over
 AdS$_2\times$S$^2$.  For various technical reasons outlined below,
 this seems to be a promising case to study. Firstly, with the
 addition of higher derivative terms one might hope to study
 corrections to the entropy beyond the Cardy limit.  It turns out that
 the near horizon geometry retains the same form upon addition of
 these higher derivative terms \cite{0701221, 0702040, 0705.1847},
 making calculations tractable.  Secondly, the rich near horizon
 geometry of the SUSY black rings allows for various approaches to
 finding the entropy to be studied and compared. In particular, the
 S$^1$ part, which carries the angular momentum of the black ring, is
 fibred over the AdS$_2$ part such that locally one has an AdS$_3$
 geometry called a self-dual orbifold of AdS$_3$ \cite{9906078,
   0311237}.  So in addition to using a generalisation of Kerr/CFT,
 corresponded to the $U(1)$ part of the isometry of near horizon geometry
 \cite{0810.2620}, the fact that one locally has an AdS$_3$ means that
 the c-extremisation formalism \cite{0506176} can be applied.  This
 approach is based on the relationship between the conformal anomaly
 and the variation of the gravitational action with respect to a
 metric of the form AdS$_3\times Y$ where $Y$ is compact. One obtains
 two Virasoro algebras corresponded to the $SL(2,{\mathbb R})_L\times
 SL(2,{\mathbb R})_R$ isometry of AdS$_3$.

 We found that comparing the result of applying Kerr/CFT and
 c-extremisation illuminating.  The fact that one only has a local
 AdS$_3$ means an application of the c-extremisation formalism is not
 straight forward. The $SL(2,{\mathbb R})_L\times SL(2,{\mathbb R})_R$
 isometry of AdS$_3$ is reduced to $SL(2,{\mathbb R})_L\times
 U(1)$. On the CFT side one expects that this kills the right-handed
 excitations, which means there is a (0,4)-CFT corresponded to SUSY
 black ring, and the microscopic entropy is given by one part of the
 Cardy formula, \cite{9711053, 0506015, 0411187, 0608012} \be
 S_{\textrm{mic}}=\sqrt{\frac{c_L\hat{q}_0}{6}}.  \ee

 On the other hand, in the Kerr/CFT approach, where the Virasoro is
 intimately related to the $U(1)$ part of the isometry, which descends from
 the right-handed $SL(2,{\mathbb R})_R$, one may naively
 expect that the central charge obtained  descends from the right-handed central charge
 of global AdS$_3$.  This is not the case. From CFT point of view, we
 know that there is a (0,4)-CFT corresponding to the SUSY black ring
 \cite{9711053, 0506015, 0411187, 0608012} and we expect that the left
 central charge contributes to the microscopic entropy.  We will show
 that both c-extremization and Kerr/CFT approaches lead to the same
 central charge which is the left central charge $c_L$.  This
 agreement was shown for supersymmetric black ring without higher
 derivative terms \cite{0810.2620}.  The equality of the two central
 charges when higher derivative corrections are added is a non-trivial new result.

 The rest of this paper is organised as follows. In section
 \ref{sec-5D} we briefly review supersymmetric black ring solutions of
 five dimensional superconformal gravity in the presence of higher
 derivative terms.  Then we apply the Kerr/CFT approach to the
 supersymmetric black ring in section \ref{sec-Kerr}. The associated
 central charge is computed and the agreement between the microscopic
 entropy and macroscopic entropy is shown. Section \ref{sec-c-extrem}
 is devoted to the application of the c-extremization formalism for
 the above black ring solution. We show that the associated left
 central charge and microscopic entropy are in agreement with the
 Kerr/CFT results.  Finally in section \ref{sec-sum}, we will
 summarize and discuss our results.

 \section{5D Supergravity with higher derivative terms}\label{sec-5D}

 In this section we review five dimensional ${\cal N}=2$ supergravity
 with higher derivative corrections associated with a mixed gauge
 gravitational Chern-Simons term.  We will do that in the context of
 an off-shell formalism which involves superconformal gravity
 \cite{0611329}.  The important feature of the formalism is that
 variation of fermionic fields does not depend explicitly on the form
 of the action.  In particular, the variation of fermionic fields does
 not change if higher order derivative terms are added.

 Compactification of M-theory on a six dimensional Calabi-Yau manifold
 results in ${\cal N}=2$ supergravity in five dimensions. In
 \cite{0611329} it was shown that by enlarging the symmetries to
 superconformal gravity (by adding some auxiliary fields) one can find
 a generic form for the fermionic variations which leave the action
 invariant. As mentioned, this form is valid for any number of higher
 derivative terms. The bosonic action, up to 4th order, is given by\footnote{We will use units $G_5 = \pi/4$ for the five dimensional Newton's constant. }
 \begin{equation}
   I=\frac{1}{16\pi G_5}\int d^5x\sqrt{|g|}
   \left(\mathcal{L}_0+\mathcal{L}_1\right),\label{action}
 \end{equation}
 where
 \begin{eqnarray}
   \mathcal{L}_0&=&\rd_a\mathcal{A}^i_\alpha\rd^a\mathcal{A}^\alpha_i+(2\nu+
   \mathcal{A}^2)\frac{D}{4}+(2\nu-3\mathcal{A}^2)\frac{R}{8}+
   (6\nu-\mathcal{A}^2)\frac{v^2}{2}+2\nu_IF^I_{ab}v^{ab}\nn\\
   &+&\frac{1}{4}\nu_{IJ}(F^{I}_{ab}F^{J\hspace{1mm}ab}+2\rd_aX^I\rd^aX^J)+
   \frac{e^{-1}}{24}C_{IJK}\epsilon^{abcde}A_a^IF^J_{bc}F^K_{de},\label{action0}
 \end{eqnarray}
 is the tree level part of the action and
 \begin{eqnarray}
   {\cal L}_1
   &=& \frac{c_{2I}}{24} \bigg( \frac{1}{16e} \epsilon_{abcde}
   A^{Ia} R^{bcfg}R^{de}_{~~fg} + \frac{1}{8}X^I C^{abcd}C_{abcd} + \frac{1}{12}X^I D^2 +
   \frac{1}{6}F^{Iab}v_{ab}D \cr
   &+& \frac{1}{3}X^I C_{abcd}v^{ab}v^{cd}+ \frac{1}{2}F^{Iab} C_{abcd}
   v^{cd} + \frac{8}{3}X^I v_{ab} \hat{\cal D}^b \hat{\cal D}_c v^{ac}
   \cr
   &+& \frac{4}{3} X^I{\hat{\cal D}}^a v^{bc} {\hat{\cal D}}_a v_{bc} +
   \frac{4}{3} X^I {\hat{\cal D}}^a v^{bc} {\hat{\cal D}}_b v_{ca}
   -\frac{2}{3e} X^I \epsilon_{abcde}v^{ab}v^{cd}{\hat{\cal D}}_f
   v^{ef}\cr
   &+& \frac{2}{3e} F^{Iab}\epsilon_{abcde}v^{cf} {\hat{\cal D}}_f
   v^{de}
   +e^{-1}F^{Iab}\epsilon_{abcde}v^c_{~f}{\hat{\cal D}}^d v^{ef}\cr
   &-& \frac{4}{3}F^{Iab}v_{ac}v^{cd}v_{db}- \frac{1}{3} F^{Iab} v_{ab}v^2 +4 X^I v_{ab}
   v^{bc}v_{cd}v^{da}-X^I (v^2)^2\bigg)\;,\label{suco}
 \end{eqnarray}
 are all four derivative terms which are related to the mixed
 gauge-gravitational Chern-Simons term $c_{2I}A^I\wedge R\wedge R$ by
 supersymmetry transformations \cite{0611329}.  In this action
 $C_{IJK}$ and $c_{2I}$ are the intersection numbers and the second
 Chern class of internal space $CY_3$ respectively,
 $\mathcal{A}^2=\mathcal{A}^i_{\alpha}\mathcal{A}_i^{\alpha}$,
 $v^2=v_{ab}v^{ab}$ and \begin{equation}
   \nu=\frac{1}{6}C_{IJK}X^IX^JX^K,\hspace{5mm}\nu_I=\frac{1}{2}C_{IJK}X^JX^K,\hspace{5mm}\nu_{IJ}=C_{IJK}X^K.
 \end{equation} The fields appearing in the action are arranged in Weyl, vector and hyper multiplets.
 The Weyl multiplet contains the metric, a 2-form auxiliary field, $v_{ab}$, a scalar auxiliary
 field $D$, a gravitino $\psi^i_\mu$ and an auxiliary Majorana spinor $\chi^i$. Each
 vector mutiplet contains a 1-form gauge field $A^I$, a scalar auxiliary field $X^I$ and a
 gaugino $\Omega^{Ii}$ (where $I=1,\cdots, n_v$ count the number of vector multiplets)  and
 $i=1,2$ is  an $SU(2)$ doublet index and
 $\alpha=1,\cdots, 2r$ refers to $USP(2r)$ group. The hyper multiplet contains the
 auxiliary scalar fields ${\cal A}^i_\alpha$ and a hyperino $\zeta^\alpha$.

 The bosonic terms of supersymmetry variation of fermions are
 \footnote{Here $\gamma_{a_1a_2\cdots
     a_m}=\frac{1}{m!}\gamma_{[a_1}\gamma_{a_2}\cdots\gamma_{a_m]}$
   which is antisymmetric in all indices. Also the covariant curvature
   $\hat{R}^{ij}_{\mu\nu}$ is defined by
   $\hat{R}^{ij}_{\mu\nu}=2\rd_{[\mu}V_{\nu]}^{ij}-2V^i_{[\mu~k}V_{\nu]}^{kj}+$
   {\em fermionic terms}, where $V_\mu^{ij}$ is a boson in the Weyl
   multiplet which is in \textbf{3} of the $SU(2)$.  For the solution
   we are going to consider, this term vanishes.}
 \begin{equation}
   \begin{array}{l}
     \delta\psi^i_\mu=\mathcal{D}_\mu\varepsilon^i+\frac{1}{2}v^{ab}\gamma_{\mu
       ab}\varepsilon^i-\gamma_\mu\eta^i,\\\\
     \delta\chi^i=D\varepsilon^i-2\gamma^c\gamma^{ab}\hat{\mathcal{D}}_av_{bc}\varepsilon^i+\gamma^{ab}\hat{R}_{ab}(V)^i_j\varepsilon^i
     -2\gamma^a\varepsilon^i\epsilon_{abcde}v^{bc}v^{de}+4\gamma^{ab}v_{ab}\eta^i,\\\\
     \delta\Omega^{Ii}=-\frac{1}{4}\gamma^{ab}F_{ab}^I\varepsilon^i-\frac{1}{2}\gamma^a\rd_aX^I\varepsilon^i-X^I\eta^i,\\\\
     \delta\zeta^\alpha=\gamma^a\rd_a\mathcal{A}^\alpha_i-\gamma^{ab}v_{ab}\varepsilon^i\mathcal{A}^\alpha_i+3\mathcal{A}^\alpha_i\eta^i,\label{sv}
   \end{array}
 \end{equation}
 where
 $\delta\equiv\bar{\epsilon}^i\textbf{Q}_i+\bar{\eta}^i\textbf{S}_i+{\xi}^a_K\textbf{K}_a$
 \footnote{$\textbf{Q}_i$ is the generator of ${\cal N}=2$
   supersymmetry, $\textbf{S}_i$ is the generator of conformal
   supersymmetry and $\textbf{K}_a$ are special conformal boost
   generators of superconformal algebra \cite{0611329}.}  and the
 covariant derivatives are defined by
 \begin{eqnarray}
   &&{\cal D}_\mu\varepsilon^i=\left(\rd_\mu+\frac{1}{4}\omega_\mu^{~ab}\gamma_{ab}+\frac{1}{2}b_\mu\right)-V_{\mu~j}^{~i}\varepsilon^j,\\
   &&\hat{\mathcal{D}}_\mu v_{ab}=\left(\mathcal{D}_\mu-b_\mu\right)v_{ab}=\rd_\mu v_{ab}+2\omega^{~c}_{[a}v_{b]c}-b_\mu
   v_{ab},
 \end{eqnarray}
 in which $b_\mu$ is a real boson in the Weyl multiplet and is $SU(2)$
 singlet \cite{0611329}.

 There is a well-known gauge to fix the conformal invariance of the
 off-shell formalism and reduce the superconformal symmetry to the
 standard symmetries of five dimensional $\mathcal{N}=2$ supergravity,
 \begin{equation}
   \mathcal{A}^2=-2,\hspace{5mm}b_\mu=0,\hspace{5mm}V_\mu^{ij}=0.\label{gauge}
 \end{equation}
 In this gauge the last equation of (\ref{sv}) gives $\eta^i$ in terms
 of $\varepsilon^i$ as,
 \begin{equation}
   \eta^i=\frac{1}{3}\gamma^{ab}v_{ab}\varepsilon^i.\label{eta}
 \end{equation}
 In the gauge (\ref{gauge}) and using above equation (\ref{eta}) the
 supersymmetry variations (\ref{sv}) simplify to
 \begin{equation}
   \begin{array}{l}
     \delta\psi^i_\mu=\left(\mathcal{D}_\mu+\frac{1}{2}v^{ab}\gamma_{\mu
         ab}-\frac{1}{3}\gamma_\mu\gamma^{ab}v_{ab}\right)\varepsilon^i,\\\\
     \delta\chi^i=\left(D-2\gamma^c\gamma^{ab}\mathcal{D}_av_{bc}
       -2\gamma^a\epsilon_{abcde}v^{bc}v^{de}+\frac{4}{3}(\gamma^{ab}v_{ab})^2\right)\varepsilon^i,\\\\
     \delta\Omega^{Ii}=\left(-\frac{1}{4}\gamma^{ab}F_{ab}^I-\frac{1}{2}\gamma^a\rd_aX^I
       -\frac{1}{3}X^I\gamma^{ab}v_{ab}\right)\varepsilon^i.
   \end{array}
   \label{1}
 \end{equation}
 In the next subsection we review the supersymmetric black ring
 solution of ${\cal N}=2$ five dimensional supergravity in the
 presence of higher derivative supersymmetric corrections
 (\ref{suco}).

 \subsection{Black Ring solution }

 To compute the entropy of an extremal black hole we just need to know
 the near horizon geometry.  In \cite{0705.1847} the near horizon of
 five dimensional supersymmetric black ring in the presence of higher
 derivative terms (\ref{suco}) is derived using the entropy function
 formalism \cite{0506177}.  In addition by using the entropy function
 formalism the macroscopic entropy of black ring is computed.
 Unfortunately it is non-trivial to extract an expression for the
 entropy purely as a function of the physical charges using this
 formalism \cite{0701221}.

 The near horizon geometry of black ring solutions with higher
 derivative terms (\ref{suco}) can be also found directly by solving
 the near horizon equations of motion \cite{0910.4907}.  In this
 subsection we report the result of these calculations and discuss the symmetries of the geometry in detail.

 The near horizon of supersymmetric black ring in five dimensions is
 given by
 \begin{eqnarray}
   &&ds^2
   =l_{AdS^2}^2\left(-r^2dt^2+{dr^2\over r^2}\right)+l^2_{S^1}\left(d\psi+e_0\,r\,dt\right)^2
   +l^2_{S^2}\left(d\theta^2+\sin^2\theta d\phi^2\right),\nn\\
   &&A^I=e^Irdt-\frac{p^I}{2}\cos\theta d\phi+a^I(e_0r\,dt+d\psi),\hspace{8mm} X^I=\frac{p^I}{l_{AdS^2}},
   \hspace{8mm}D=\frac{12}{{l^2_{AdS^2}}},\label{entropy-function}\\
   &&Q^I=-4C_{IJK}p^Ja^K,\hspace{15mm}e^I+e_0a^I=0,\hspace{15mm}v_{\theta\phi}=\frac{3}{8}{l_{AdS^2}}\sin\theta,\nn
 \end{eqnarray}
 in which $\theta$ and $\phi$ are the coordinates of a usual 2-sphere
 and $\psi$ is the coordinate of ring and is periodic,
 $\psi\sim\psi+4\pi$, $Q^I$ are the electric charges and the radii are given by the magnetic charges
 $p^I$,
 \begin{equation}
   l_{AdS_2}=l_{S^2}=e_0\,l_{S^1} ={1\over2}\left({1\over6}C_{IJK}p^Ip^Jp^K+{1\over12}c_{2I}p^I\right)^{1/3}.
 \end{equation}
 From (\ref{entropy-function}) we can see that the metric consists of
 a $U(1)$ fibred over an $AdS_2$ base times a two-sphere. In other
 words the isometries of the metric are $SL(2,{\mathbb R})\times
 U(1)\times SO(3)$ and are generated by
 \begin{eqnarray}
   L_0&=&r\rd_r-t\rd_t,\quad L_1=(t^2+r^{-2})\rd_t-2rt\rd_r-\frac{2e_0}{r}\rd_\psi,\quad
   L_{-1}=\rd_t,\nn \\
   \bar{L}_0&=&-e_0\rd_\psi,\label{isometry}\\
   J^3&=&-i\rd_\phi,\quad
   J^\pm=e^{\pm i\phi}(-i\rd_\theta\pm\cot\theta\rd_\phi).\nn
 \end{eqnarray}

 In fact we can think of the first part of the metric as locally $AdS_3$ with the
 symmetries $SL(2,{\mathbb R})\times SL(2,{\mathbb R})$ with the one
 of the $ SL(2,{\mathbb R})$'s broken to a $U(1)$.  One can show that
 we also have the locally defined killing vectors
 \begin{eqnarray}
 \bar{L}_1=e^{\frac{\psi}{e_0}}\left(\frac{1}{r}\rd_t+r\rd_r-e_0\rd_\psi\right),\hspace{12mm}
 \bar{L}_{-1}=e^{-\frac{\psi}{e_0}}\left(\frac{1}{r}\rd_t-r\rd_r-e_0\rd_\psi\right).
 \end{eqnarray}
 Notice that since $\psi$ is periodic, $\bar{L}_{\pm 1}$ are not
 well defined globally.
 One finds that together with the generator of the $U(1)$
 part of the near horizon isometry, $\bar{L}_0$, we obtain an $SL(2,
 \mathbb{R})$ algebra,
\begin{equation}
   [\bar{L}_m,\bar{L}_n]=(m-n)\bar{L}_{m+n},\hspace{10mm}m,n=0, \pm1.
\end{equation}
  The periodicity of $\psi$ breaks this $SL(2, \mathbb{R})$ to a
 $U(1)$.  This local $AdS_3$ symmetry  will permit us to use the
 c-extremization approach \cite{0506176} to find the associated
 central charge in section \ref{sec-c-extrem}.

Now, the parameter $e_0$ gives the angular momentum of the black ring
 solution in 5D, while if one reduces along the $\psi$ direction, from
 a 4D point of view, $e_0$ is an electric field.  In the entropy function
 formalism one can not easily find this electric field in terms of physical
 charges (or angular velocity) of the black ring \cite{0705.1847,
   0810.2620}.

 Solving the near horizon equations of motion leads to an additional
 relation between the parameter $e_0$, magnetic charges $p^I$ of the
 black ring, electric charges $Q^I$ of the black ring and angular
 velocities $J_\phi$ and $J_{\psi}$,
 \begin{eqnarray}
   J_\phi-J_\psi+{1\over8}C^{IJ}(Q_I-C_{IK}p^K)(Q_J-C_{JL}p^L)&=&{1\over {e_0}^2}\left({1\over6}C_{IJK}p^Ip^Jp^K+{1\over6}c_{2I}p^I\right).\label{e-def}\\
   J_\phi&=&{1\over2}p^I\left(Q_I-{1\over36}C_{IJK}p^Jp^K\right),
 \end{eqnarray}
 where $C^{IJ}$ is the inverse of $C_{IJ}\equiv C_{IJK}p^K$.  In
 \cite{0910.4907}, it was also shown that the left hand side of
 eq.(\ref{e-def}) is equal to the charge $\hat{q}_0$, associated with
 the Kaluza-Klein photon, which will be used to compute the
 microscopic entropy of black ring in the next section,
 \begin{equation}
   \hat{q}_0\equiv J_\phi-J_\psi+{1\over8}C^{IJ}(Q_I-C_{IK}p^K)(Q_J-C_{JL}p^L).\label{q0}
 \end{equation}

 The macroscopic entropy of supersymmetric five dimensional black ring
 is given by
 \begin{equation}
   S_{\textrm{mac}}={2\pi\over e_0}\left({1\over6}C_{IJK}p^Ip^Jp^K+{1\over6}c_{2I}p^I\right)
   =2\pi\sqrt{{{\hat{q}_0(C_{IJK}p^Ip^Jp^K+c_{2I}p^I)}\over6}}.\label{mac}
 \end{equation}
 In the next two sections we will compute the microscopic entropy of
 black rings with higher derivative corrections (\ref{suco}).  We will
 show that both formalisms lead to the same result for the microscopic
 entropy and the macroscopic entropy calculated in this section
 (\ref{mac}).


 \section{Kerr/CFT Approach}\label{sec-Kerr}
 The microscopic entropy of extremal black rings can calculated by
 using the Kerr/CFT approach.  This approach can be applied when the
 near horizon geometry contains a $U(1)$ fibred over AdS$_2$ which is
 the case for black rings we consider.

 The Kerr/CFT approach was extended to the case with a Chern-Simons
 term \cite{0902.1001}.  It was shown that for a theory with gravity
 and also other fields, the central charge is not affected by
 non-gravitational fields.  This approach was also generalized to
 theories with higher derivative corrections \cite{0903.4176}.
 Although this generalization was based on four dimensional kerr black
 hole in the extremal limit we will show that the black ring satisfy
 the conditions that help us to use the results of \cite{0903.4176} to
 compute the central charge of associated Virasoro algebra in the
 presence of higher derivative corrections (\ref{suco}).
 \subsection{Asymptotic symmetry group}
 Since the black ring near horizon geometry with higher derivative
 corrections is similar to the case without, one can use the same
 boundary conditions as those used in \cite{0810.2620},
 \begin{equation}
   h_{\mu\nu}\sim\mathcal{O}\left(\begin{array}{ccccc}
       r^2&~ 1/r^2&~ 1/{r}&~ r &~ 1 \\
       ~&~{1}/{r^3}&~ {1}/{r^2}&~{1}/{r}^2&~{1}/{r}\\
       ~&~&~{1}/{r}&~{1}/{r}&~{1}/{r}\\
       ~&~&~&~1/r&~1\\
       ~&~&~&~&~1
     \end{array}\right),\label{h}
 \end{equation}
 in the basis $(t, r, \theta, \phi, \psi)$. The generators
 associated to these boundary conditions are given by
 \begin{equation}
   \zeta_n=-e^{-in\psi}\rd_{\psi}-in~r~e^{-in\psi}\rd_r,\label{ln}
 \end{equation}
 which satisfy a Virasoro algebra
 \begin{equation}
   i[\zeta_m,\zeta_n]=(m-n)\zeta_{m+n}.\label{vira}
 \end{equation}
 Two interesting facts can be noted when comparing (\ref{isometry})
 and (\ref{ln}).  Firstly, $\zeta_0$ is proportional to $\bar L_0$ which is
 the generator of the near horizon $U(1)$ symmetry. It is said that
 the Virasoro is ``based'' on this $U(1)$.  Secondly, the other $\zeta$'s
 do not commute with $L_1$ which is a generator of the near horizon
 $SL(2,{\mathbb R})$. Furthermore, this non-commutativity is due to
 the last term of $L_1$ which is related to the fibration of the
 $U(1)$ on an $AdS_2$ base. This means that the Virasoro is not
 decoupled from the $SL(2,{\mathbb R})$.

 To apply the Kerr/CFT approach when higher derivative corrections are
 added it is useful to do the calculations in a non-basis coordinates.
 The vielbeins associated to near horizon geometry of black ring are
 \begin{equation}
   e^{\hat{t}}=l_{AdS_2}r dt,\hspace{3mm}e^{\hat{r}}=\frac{l_{AdS_2}}{r}dr,\hspace{3mm}
   e^{\hat{\theta}}=l_{AdS_2}d\theta,\hspace{3mm}e^{\hat{\phi}}={l_{AdS_2}\sin\theta}d\phi,\hspace{3mm}
   e^{\hat{\psi}}=l_{S^1}(d\psi+{e_0}r\,dt),\label{vielbein}
 \end{equation}
 and the variations of the veilbeins are given by
 \begin{eqnarray}
   &&\mathcal{L}_{\zeta_n}e^{\hat{t}}=i\,n\,e^{-in\psi}e^{\hat{t}},\hspace{30mm}
   \mathcal{L}_{\zeta_n}e^{\hat{r}}=-{e_0}\,n^2\,e^{-in\psi}\left(e^{\hat{\psi}}-e^{\hat{t}}\right),\nn\\
   &&\mathcal{L}_{\zeta_n}e^{\hat{\theta}}=\mathcal{L}_{\zeta_n}e^{\hat{\phi}}=0,\hspace{29mm}
   \mathcal{L}_{\zeta_n}e^{\hat{\psi}}=i\,n\,e^{-in\psi}\left(e^{\hat{\psi}}-2e^{\hat{t}}\right).\label{lv}
 \end{eqnarray}
 These variations are similar to the case of the Kerr black hole
 \cite{0903.4176}.

 The Virasoro algebra we found (\ref{vira}) corresponds to Poisson
 brackets between the generators. Since we are interested in studying
 the quantum behavior of the boundary fluctuations, we need to find
 the Dirac brackets which may lead to a Virasoro algebra with a
 central charge.  To compute this central charge we follow
 \cite{0111246, 0708.2378, 0708.3153}.  The central charge is given by
 \begin{equation}
   c^{(k)}=12\,i\,\int_{\rd\Sigma}\emph{\textbf{k}}^{inv}_{\zeta_n}[\mathcal{L}_{\zeta_{-n}}g;g]\,\bigg|_{n^3}
 \end{equation}
 where $|_{n^3}$ stands for the term of order $n^3$ and
 \begin{equation}
   \emph{\textbf{k}}^{inv}_{\zeta_n}[\mathcal{L}_{\zeta_{-n}}g;g]=
   -2\left[\emph{{\textbf X}}_{cd}\mathcal{L}_{\zeta_{n}}\nabla^c\zeta^d_{-n}+
     (\mathcal{L}_{\zeta_n}\emph{{\textbf X}})_{cd}\nabla^{[c}\zeta^{d]}_{-n}+
     \mathcal{L}_{\zeta_n}\emph{{\textbf W}}_c\zeta^c_{-n}\right]
   -\emph{{\textbf E}}[\mathcal{L}_{\zeta_{n}}g,\mathcal{L}_{\zeta_{n}}g;g],
 \end{equation}
 in which covariant derivatives are defined with respect to the
 original metric $g$.  $\emph{{\textbf X}}$ and $\emph{{\textbf W}}$
 are related to $Z^{abcd}$, the variation of the Lagrangian with
 respect to the Riemann tensor $R_{abcd}$,
 \begin{equation}
   Z^{abcd}=\frac{\delta^{cov}L}{\delta R_{abcd}},
 \end{equation}
 by,
 \begin{equation}
   (\emph{{\textbf W}}^c)_{c_3c_4c_5}=-2\nabla_dZ^{abcd}\epsilon_{abc_3c_4c_5}
   =2(\nabla_d\emph{{\textbf X}}^{cd})_{c_3c_4c_5}.
 \end{equation}

 The process of finding the central charge for the supersymmetric
 black ring follows the same recipe as for the Kerr solution.
  After some work one finds that the central charge associated to the
 Virasoro algebra (\ref{vira}) is,
 \begin{equation}
   c^{(k)}=-12{e_0}\int_\Sigma Z_{abcd}\epsilon^{ab}\epsilon^{cd}\textrm{vol}(\Sigma)
   ={6\,{e_0}\over\pi}\,S_{\textrm{mac}}.
 \end{equation}
 In the last step we used the Iyer-Wald formula for macroscopic
 entropy of a black hole which is generalization of Bekenstein-Hawking
 formula when the higher derivative terms are appeared.  So the cental
 charge is
 \begin{equation}
   c^{(k)}=C_{IJK}p^Ip^Jp^K + c_{2I}p^I.\label{c-k}
 \end{equation}
 As we shall see in the next section, this central charge is equal to
 the left central charge computed by the c-extremization formalism.
 This equality was shown for black rings without higher derivative
 corrections in \cite{0810.2620}.  Finding this relation for the case
 with higher derivative terms is a much stronger result and unlikely
 to be a coincidence.  We consider this equality further in the
 discussion section.
 \subsection{Microscopic entropy}

 The microscopic entropy of supersymmetric black ring in the Kerr/CFT
 approach can be computed by the following form of the Cardy formula,
 \begin{eqnarray}
   S_{\textrm{mic}}^{(k)}={\pi^2\over3}\,c^{(k)}\,T_{FT},\label{cardy2}
 \end{eqnarray}
 where $T_{FT}$ is the Frolov-Thorne temperature. The Frolov-Thorne
 temperature is an intrinsic feature of metric and its definition is
 not corrected by higher derivative terms.\footnote{Appendix B of
   \cite{0903.4176} is devoted to this subject.}  So as usual, one can
 find the Frolov-Thorne temperature from the $t\psi$ cross term of
 near horizon geometry (\ref{vielbein})\footnote{Often there is a
   factor of 2 in the denominator of the expression for the
   Frolov-Thorne temperature but not in our case since we have take
   the period of $\psi$ to be $4\pi$.}
 \begin{equation}
   T_{\textrm{FT}}={1\over\pi{e_0}}.\label{tft}
 \end{equation}

 Using (\ref{cardy2},\ref{tft}) one finds that
 \begin{eqnarray}
   S_{\textrm{mic}}^{(k)}=
   {2\pi\over {e_0}}\left({1\over6}C_{IJK}p^Ip^Jp^K+{1\over6}c_{2I}p^I\right).\label{mic2}
 \end{eqnarray}

 As we expect this microscopic entropy associated with the asymptotic
 symmetry group is equal to the macroscopic entropy (\ref{mac}).


 \section{C-extremization approach}\label{sec-c-extrem}

 One can also use the usual Cardy formula to compute the microscopic
 entropy of the supersymmetric black ring.
 The low energy decoupling limit implies that only
 the left-handed excitations survive and the microscopic entropy is given by,
 \begin{equation}
   S_{\textrm{mic}}=2\pi\,\sqrt{{c_L\,\hat{q}_0\over6}}.\label{cardy1}
 \end{equation}
 In this form of the Cardy formula $c_L$ is the left central charge
 and $\hat{q}_0$, given in eq.(\ref{q0}), corresponds to the
 left-handed excitations of the CFT .  Using the c-extremization
 formalism one can compute this central charge from near horizon data.
 Although this formalism was introduced for a geometry with a globally
 AdS$_3$ part, we assume that it can also be used for geometries which
 are locally AdS$_3$.  We don't prove this but the self-dual orbifold
 AdS$_3$ perspective \cite{9906078, 0311237} and EVH/CFT proposal \cite{1107.5705} suggest that this
 approach can also be used for a locally AdS$_3$ geometry. A
 posteriori the fact that the we obtain non-trivial agreement with the
 results of the previous section is gives further weight to our
 assumption.  We will discuss this point in last section.

 At the leading order it was shown that the c-extremization and
 Brown-Henneaux (or Kerr/CFT) approach lead to the same result for the
 central charge \cite{0810.2620}.  At this level the left and right
 central charges are equal and given by
 \begin{equation}
   c_L=c_R=C_{IJK}p^Ip^Jp^K.
 \end{equation}
 Turning on the higher order correction one can use the
 c-extremization approach to find the average of left and right
 central charges.  Then, finding the gravitational anomaly gives the
 difference between left and right central charges so that combining
 the two one can obtain $c_L$ and $c_R$.

 The first step in applying the c-extremization formalism is choosing
 an appropriate ansatz,
 \begin{eqnarray}
   &&ds^2=l^2_{AdS_3}ds^2_{AdS_3}+l^2_{S^2}ds^2_{S^2},\\
   &&A^I=e^Irdt-\frac{p^I}{2}\cos\theta d\phi+a^I(e_0r\,dt+d\psi),\hspace{8mm}
 \end{eqnarray}
 Then by extremizing the c-function,
 \begin{equation}
   c=6{l_{AdS_3}}^3{l_{S^2}}^2({\cal L}_0+{\cal L}_1),
 \end{equation}
 with respect to, ${l_{AdS_3}}$ and $l_{S^2}$, the AdS and sphere
 radii respectively, we find their values in terms of the magnetic
 charges.  The value of c-function at these radii gives the average of
 left and right central charges.  Performing these calculation one
 finds,
 \begin{eqnarray}
   &&l_{AdS_3}=2\,l_{S^2} =\left({1\over6}C_{IJK}p^Ip^Jp^K+{1\over12}c_{2I}p^I\right)^{1/3},\\
   &&A^I=e^Irdt-\frac{p^I}{2}\cos\theta d\phi+a^I(e_0r\,dt+d\psi),\hspace{8mm} X^I=\frac{p^I}{l_{AdS^2}},
   \hspace{8mm}D=\frac{12}{{l^2_{AdS^2}}},\\
   &&Q^I=-4C_{IJK}p^Ja^K,\hspace{15mm}e^I+e_0a^I=0,\hspace{15mm}v_{\theta\phi}=\frac{3}{8}{l_{AdS^2}}\sin\theta,\nn
 \end{eqnarray}
 and the value of c-function at this extremum point is given by
 \begin{equation}
   c|_{ext.}={1\over2}(c_L+c_R)=C_{IJK}p^Ip^Jp^K+{3\over4}c_{2I}p^I.
 \end{equation}
 There is a precise agreement between the above solution and the
 results of entropy function formalism reviewed in the previous
 section (\ref{entropy-function}).

 In \cite{0506176} it was shown that for the associated dual CFT the
 gravitational anomaly yields the difference between left and right
 central charges,
 \begin{equation}
   c_L-c_R={1\over2}c_{2I}p^I.
 \end{equation}
 Thus the left and right central charges are given by
 \begin{equation}
   c_L =C_{IJK}p^Ip^Jp^K + c_{2I}p^I ,\hspace{10mm} c_R =C_{IJK}p^Ip^Jp^K +\frac{1}{2} c_{2I}p^I ,\label{c-c}
 \end{equation}

 Now we can use the Cardy formula (\ref{cardy1}) and equations
 (\ref{q0}) and (\ref{e-def}) to compute the microscopic entropy of
 black ring,

 \begin{eqnarray}
   S_{\textrm{mic}}^{(c)}=2\pi\sqrt{{c_L\,\hat{q}_0\over6}}=
   {2\pi\over e_0}\left({1\over6}C_{IJK}p^Ip^Jp^K+{1\over6}c_{2I}p^I\right).\label{mic1}
 \end{eqnarray}
 The above entropy is in precise agreement with result of Kerr/CFT
 approach (\ref{mic2}).


 \section{Summary and discussion}\label{sec-sum}

 In this paper we study the microscopic interpretation of SUSY black
 ring solutions of ${\cal N}=2$ supergravity in the presence of
 supersymmetric completion of mixed gauge-gravitational Chern-Simons
 term (\ref{suco}). Because of the near horizon geometry of these
 solutions, one can use both c-extremization and the Kerr/CFT approach to
 find the microscopic entropy via computing the associated central
 charge. We showed that central charge, which counts the degeneracy of
 ground states in the CFT side, is given by the magnetic charges of
 SUSY black rings (\ref{c-c}) or (\ref{c-k}) in both methods
 independently.

 We found that the usual form of the Cardy formula (\ref{cardy1})
 without any subleading corrections can be used for SUSY black ring
 solution even in the presence of higher derivative corrections
 (\ref{suco}). \footnote{This was previously shown for BTZ black hole
   solutions of 3D gravity \cite{gr-qc/9909061}. } This works because
 the effect of the higher derivative corrections is relatively simple
 -- essentially we just have a shift of the Kaluza-Klein photon charge
 (\ref{e-def},\ref{q0}). As long as we remain in the large charge
 regime, we do not need to consider subleading corrections. It means
 our results are in agreement with the canonical ensemble description
 of black ring used in \cite{1010.3561}.  This discussion also applies
 to the use of (\ref{cardy2}).

 The most interesting result of our study is that the Kerr/CFT and
 c-extremization approaches which are apparently related to different
 Virasoro algebras lead to the same value of central charges and the
 same microscopic entropies in a highly non-trivial setting.

 In order compare these approaches, and to try get a handle on the
 various central charges appearing in the game, it is helpful to
 consider the geometry in detail.  While global $AdS_3$ has two
 $SL(2,{\mathbb R})$ symmetries, as discussed in section \ref{sec-5D},
 our near horizon only has one with the other $SL(2,{\mathbb R})$ is
 broken to a $U(1)$. Now, it would seem to be natural to associate
 $c_L$ with the unbroken $SL(2, \mathbb{R})$ and conclude that the
 right-handed excitations are killed by the broken $SL(2,
 \mathbb{R})$.  This however seems to be incompatible with the fact
 that it is precisely the residual part of the broken $SL(2,
 \mathbb{R})$, $\bar{L}_0$ which forms the basis of the Virasoro
 algebra associated with the asymptotic symmetry group considered in
 the Kerr/CFT approach.  The incompatibility is however not so stark
 once one realises that c-extremization focuses on the near horizon
 geometry of the solutions, while the Kerr/CFT approach is based on
 the fluctuations at the boundary (of the near horizon geometry). Our results are strong evidence
 that both approaches count microstates of the same CFT but in
 different ways.  In fact, the generators of the unbroken $SL(2,
 \mathbb{R})$, $L_0, L_{\pm1}$, do not commute with the Virasoro
 generators of the Kerr/CFT approach (\ref{ln}) which means they are
 not independent. This suggests that although the asymptotic symmetry
 group Virasoro algebra is ``based'' on the $U(1)$ part of the near
 horizon isometry, the $SL(2,\mathbb{R})$ plays a crucial rule in
 Kerr/CFT correspondence.

 This leads us to conclude that both points of view are somehow talking to each
 other.  The central charge counts the ground
 states dual to SUSY black rings and the number of these states is
 independent of any approach used to this counting.

 This is in agreement with the DLCQ approach
 which relates the left-handed excitations of the self-dual orbifold
 AdS$_3$ geometry on two distinct boundaries \cite{0906.3272,
   1011.1803, 1011.1897}. In our case, The S$^1$ part, which carries the angular momentum of the black ring,
 is fibred over the AdS$_2$ part such that locally one has an AdS$_3$
 geometry called a self-dual orbifold of AdS$_3$ \cite{9906078,
   0311237}.  From this viewpoint it was shown that extremal BTZ black
 hole, which has self-dual orbifold AdS$_3$ near horizon geometry, is
 dual to discrete light cone quantised CFT$_2$ which admits one chiral
 Virasoro algebra \cite{0906.3272, 1011.1803, 1011.1897}.

 Our results suggest some further possible avenues to investigate.
 It would be interesting to compare three dimensional extremal BTZ
 black holes in the presence of some higher derivative terms and   light cone
 quantised CFT$_2$.  Another interesting avenue
to consider is whether hidden conformal symmetries appear beyond the extremal limit
for SUSY black rings with higher derivative corrections. In this situation one expects
both left and right central charges to be excited.\\

 {\large Acknowledgements}\\
 We would like to thank M. M. Sheikh-Jabbari for many helpful comments
 and insights on a preliminary draft of this work and Robert de Mello
 Koch and Vishnu Jejjala for useful discussion.  This work is based upon research
 supported by National Research Foundation. Any opinion, findings and
 conclusions or recommendations expressed in this material are those
 of the authors and therefore the NRF do not accept any liability with
 regard thereto.


\bibliographystyle{utphys}
\bibliography{BR-CFT}

 \end{document}